# Raman fiber amplifier with integrated cooler

**Galina Nemova[1,*] and Raman Kashyap[1,2]**

*Department of Engineering Physics[1] and Department of Electrical Engineering[2],*

*École Polytechnique de Montréal, P.O. Box 6079, Station Centre-ville, Montréal, Canada*

[*]*Corresponding author: galina.nemova@polymtl.ca*

**Abstract:** We present a new scheme for a laser cooled Raman optical fiber amplifier. The heat generated in the process of stimulated Raman scattering is compensated for with laser cooling provided by anti-Stokes fluorescence of ytterbium ions doped in the core of the fiber. The device is pumped with two pump sources. One of the pump sources provides Raman amplification of the signal and does not interact with the ytterbium ions in the core of the fiber. The second pump provides laser cooling of the fiber with anti-Stokes fluorescence of the $Yb^{3+}$ ions. The proper arrangement of $Yb^{3+}$ distribution can provide athermal performance of the device.

## 1. Introduction

Stimulated Raman scattering (SRS) in optical fibers is a well-known and widely used method of amplification and generating coherent signals at a wide range of wavelengths. Amplifiers and lasers designed for telecommunications applications in the wavelength range 1-1.5 *μm* are based on silica glass fibers. Recently, the wavelength range beyond 2 *μm* has attracted attention for a number of applications in bio-photonics. Nonoxide glass fibers with low maximum phonon energies and high infrared transparency such as fluoride, tellurite, and chalcogenide are the choices for this wavelength range. An important issue in high power Raman fiber amplifiers is the heat generated inside the active medium due to the quantum defect between the pump and the amplified wavelengths. The increase in the temperature of the laser medium can detrimentally affect laser performance, by degrading the quality of the amplified beam, causing a decrease in the efficiency of the process, thermal lensing and self focussing. Heat loading can change the gain or even damage the material of the amplifier. The cooling of the amplifier medium is a key issue for improving its performance and increasing the amplifier's lifetime. Air-cooling of fibers is inefficient for high powers and water cooling should be avoided especially in the case limited chemical stability such as with fluoride fibers. On the other hand, any surface cooling results in a thermal gradient that strains the laser medium and distorts optical waves. Waste heat is a major limiting factor in scaling lasers and amplifiers.

In this paper we suggest to use laser cooling of solids to compensate for the heat generated in the SRS process, for the first time of our knowledge. Laser cooling of the fiber is provided with anti-Stokes fluorescence of $Yb^{3+}$ ions doped in the core of the fiber, which are not involved in the amplification process, but can provide local heat compensation, free from any temperature gradient. The optically pumped $Yb^{3+}$ions in the core of the fiber can be defined as an *integrated optical cooler*. A theoretical description of the amplifier performance is presented in Section 2. The results of the simulation are discussed in detail in Section 3.

## 2. Theoretical Analysis

The concept of using anti-Stokes fluorescence to cool solid state matter was first proposed by Peter Pringsheim in 1929 [1]. It has been shown that some materials emit light at shorter wavelengths than the illuminating wavelength due to thermal (phonon) interactions with the excited atoms [2]. This process is called anti-Stokes fluorescence as opposed to Stokes fluorescence in which the emitted wavelength is longer than that absorbed one. Since anti-Stokes fluorescence involves the emission of higher energy photons than those which are absorbed, the net anti-Stokes fluorescence can remove energy from the material, and as a consequence, lead to its refrigeration. Net radiation cooling by anti-Stokes fluorescence in solid materials was observed experimentally for the first time in 1995 by Epstein's research team in $Yb^{3+}$-doped fluorozirconate ZBLAN glass [3]. More recently, laser cooling of solids was observed in different $Yb^{3+}$-doped low phonon materials. The $Yb^{3+}$ ions have two manifolds and as a consequence are free from excited-state absorption (ESA), which adversely affects the cooling process. Ytterbium ions are the most widely used ions in experiments devoted to laser cooling of solids [4-7] although laser induced cooling with other rare-earth (RE) ions such as $Tm^{3+}$ and $Er^{3+}$ has been also reported [8, 9].

In this paper, a high power optical fiber Raman amplifier based on ZBLAN glass with $Yb^{3+}$-doped core, in which $Yb^{3+}$ ions serve as an integrated cooler and do not participate in the amplification process, is presented. The scheme of the laser cooled Raman fiber amplifier is presented in Fig. 1. It has two pump sources. The pump beam with the power, $P_p$, propagates at the wavelength $\lambda_p = 2.5\ \mu m$. It is responsible for the SRS process and co-propagates with the amplified signal, $P_s$, propagating at the wavelength $\lambda_s = 2.575\ \mu m$, equal approximately to the the first Stokes wavelength. The core of the fiber is doped with $Yb^{3+}$ ions. The concentration of the ions changes non-uniformly along the length of the fiber. It is illustrated by the curves in the core of the fiber (Fig.1). A second pump beam called the cooling pump beam propagates with the power, $P_p^{cool}$, at the wavelength $\lambda_p^{cool} = 1.015$ $\mu m$, which is longer than the effective mean fluorescence wavelength ($\lambda_{f^*} = 0.9976$ $\mu m$) of the $Yb^{3+}$ ions in ZBLAN. It excites the ions to provide anti-Stokes fluorescence causing optical cooling. The anti-Stokes fluorescence leaving the structure is illustrated by the dashed arrows in Fig. 1. This anti-Stokes fluorescence can propagate along the length of the fiber as cladding modes being re-absorbed, deteriorating the cooling process. The roughness of the outer cladding surface of the fiber can be used to cause scattering of the cladding modes and prevent the re-absorption.

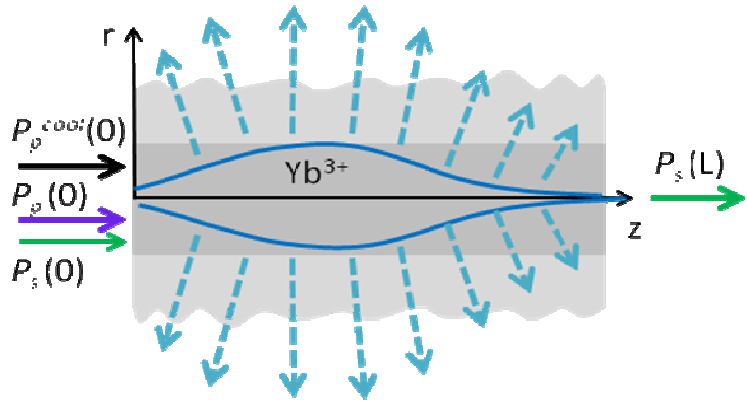

Fig. 1. Structure under consideration. The curves in the fiber core illustrate the distribution of $Yb^{3+}$ ions. The dashed arrows illustrate anti-Stokes fluorescence.

*A. Temperature distribution in an un-cooled Raman amplifier*

To appreciate the significance of the suggested scheme the distribution of the increase in the temperature of the fiber along the length of the fiber starting from the room temperature caused by the Raman amplification process when cooling is not applied, is now considered. The change in the power of the pump and the signal beams along the length of the fiber can been calculated with the equations describing the Raman amplification process [10]

$$\frac{\partial P_S}{\partial z} = \chi_R^{(3)} \kappa_{ps} P_s P_p - \alpha_s P_s,$$
$$\frac{\partial P_p}{\partial z} = -\frac{\omega_p}{\omega_s} \chi_R^{(3)} \kappa_{ps} P_p P_s - \alpha_p P_p, \tag{1}$$

where
$$\kappa_{ps} = 2\pi \frac{\omega_s}{cZ_0} (A^p)^2 (A^s)^2 \int_0^a J_0^2(\kappa_{01}^p r) J_0^2(\kappa_{01}^s r) r dr. \tag{2}$$

Here $\chi_R^{(3)}$ is the Raman susceptibility, $\omega_p$ and $\omega_s$ are the frequencies of the pump and Stokes modes, respectively. $c$ is the speed of light in vacuum, $Z_0 \approx 377\ \Omega$ is the intrinsic impedance. $A^p$ and $A^s$ are the amplitudes of the fiber core modes supporting the pump and Stokes modes, respectively.

$$A^{p,s} = \frac{\gamma^{p,s}}{a}\left[\frac{Z_0}{\pi n_{eff}^{p,s}\left(n_{co}^2 - n_{cl}^2\right)J_1^2\left(\kappa_{01}^{p,s}a\right)}\right]^{1/2}, \qquad (3)$$

where $a$ is the radius of the core of the fiber, $n_{co}$ and $n_{cl}$ are refractive index of the fiber core and cladding, respectively, $n_{eff}^{p,s}$ are effective refractive index of the core mode at the wavelengths $\lambda_p$ and $\lambda_s$, respectively:

$$\gamma^{p,s} = \sqrt{\left(n_{eff}^{p,s}\right)^2 - n_{co}^2}, \quad \text{and} \quad \kappa_{01}^{p,s} = \sqrt{n_{co}^2 - \left(n_{eff}^{p,s}\right)^2}. \qquad (4)$$

Throughout the amplification process each pump photon, which is not Rayleigh scattered, generates a Stokes photon with lower energy than the pump photon. The rest of the energy of the pump photon causes vibrations of the lattice leading to a rise in the temperature of the fiber. As is seen from the theoretical estimations the number of the Rayleigh scattered pump photons is unimportant in comparison with the number of photons participating in the SRS process if the pump beam propagates at the wavelength $\lambda_p = 2.5\ \mu m$ corresponding to a minimum of the fiber loss. The heat density generated during the SRS process in the core of the fiber can be estimated as

$$I_{heat} = \frac{dP_p}{dz}\left(1 - \frac{\lambda_p}{\lambda_s}\right)\frac{1}{A_{eff}}, \qquad (5)$$

where $A_{eff}$ is effective mode area of the fiber core mode. Thermal effects in the fiber under deposition of heat power have been discussed in Ref. 11. The average temperature of the fiber can be calculated using the formula [11]:

$$T_{av} = T_c + \frac{I_{heat}a^2}{2bH} - \frac{I_{heat}a^3}{3b\kappa}, \qquad (6)$$

where $T_c$ is the room temperature, $a$ and $b$ are the radii of the fiber core and cladding, respectively. Here $\kappa$ denotes the thermal conductivity. $H$ is the convective coefficient.

*B. Laser cooling in the RE-doped glass*

During the cooling cycle with RE ions, the pump photons excite ions from the top of the ground state to the bottom of the excited state. The excited ions, absorbing phonons during thermalization process reach quasi-equilibrium with the lattice. Broadband anti-Stokes fluorescence follows with an effective mean photon energy of $hc/\lambda_{f^*}$, where $h \approx 6.626 \times 10^{-34}$ J·s is Planck's constant, $\lambda_{f^*} = [1/\lambda_f - \kappa_{nr}/(hc\gamma_{rad})]^{-1}$. Here $\kappa_{nr}$ quantifies non-radiative processes re-heating the sample; $\lambda_f$ can be estimated using the formula

$$\lambda_f = \frac{\int I_f(\lambda)\lambda d\lambda}{\int I_f(\lambda)d\lambda}, \qquad (7)$$

where $I_f(\lambda)$ is the fluorescence spectral intensity. The cooling intensity at any point inside the sample is equal to the difference between the spontaneously radiated intensity and absorbed pump intensity and can be calculated using the expression [12]

$$I_{cool} = \frac{N(z)\sigma_{abs}(\lambda_p^{cool})P_s(\lambda_p^{cool} - \lambda_{f^*})}{\lambda_{f^*}\left(1 + \frac{\sigma_e(\lambda_p^{cool})}{\sigma_{abs}(\lambda_p^{cool})} + \frac{P_s}{P_p}\right)A_{eff}}, \qquad (8)$$

where $I_s = hc\gamma_{rad}/[\lambda_p\sigma_{abs}(\lambda_p{}^{cool})]$. Here $\sigma_{abs}(\lambda)$ and $\sigma_e(\lambda)$ are the absorption and stimulated-emission cross sections at the wavelength $\lambda$, respectively. $\gamma_{rad}$ is the radiative relaxation rate of the excited-state manifold at room temperature. $N(z)$ is the density of the $Yb^{3+}$ ions distributed along the length of the fiber.

## 3. Results and discussion

In this part of the paper computer simulations describing the performance of the proposed scheme are presented. In the simulations a fiber is considered, which can support single mode operation of the pump, $P_p$, as well as the signal, $P_s$, beams. The structure with radii of the core $a = 10\ \mu m$ and cladding $b = 50\ \mu m$, respectively is assumed. The refractive index of the core is $n_{co} = 1.5$, and the refractive index difference with the cladding is $12\times10^{-3}$. The distribution of the deviation of the average temperature of the fiber from room temperature ($\Delta T = T_{av} - T_c$) along the length of the fiber when cooling is not applied, is calculated with equation (6), and illustrated in Fig. 2 for two input pump powers $P_p(0) = 80$ W and $P_p(0) = 100$ W. The signal power at the input to the amplifier $P_s(0) = 5$ W. The distribution of the deviation of the average temperature of the fiber from room temperature has a maximum, which corresponds to the position of the intense amplification of the signal, accompanied by a tremendous increase in the number of the phonons ("superfluous" phonons) causing localized heating of the fiber amplifier. The evolution of the power of the pump and amplified beams along the length of the fiber has been calculated with Eqs. (1) – (4) and illustrated in Fig. 3. For very high powers this increase in the temperature of the fiber can cause unwanted effects or even damage the fiber. Laser cooling with RE ions doped in the active medium of the amplifier can prevent this from happening. Indeed the "superfluous" phonons generated in the SRS process and causing heating can be removed by the cooling with anti-Stokes fluorescence of $Yb^{3+}$ ions doped in the core of the fiber. If all "superfluous" phonons generated in the lattice throughout Raman amplification process will be removed with the laser cooling process the temperature of the fiber will remain constant and equal to the temperature of the ambient environment. Equating expressions (8) and (5) one can calculate the distribution of the $Yb^{3+}$ ions in the core of the fiber, which can completely compensate for the heat generated in the SRS process. Such distributions of the ions for two input pump powers $P_p(0) = 80$ W and $P_p(0) = 100$ W are illustrated in Fig. 4. They are non-uniform and have the maximum of ion concentration, which corresponds to the maximum intensity of the phonons generated in the fiber core. The cooling pump power, $P_p^{cool}$, is below the Raman and Brillouin thresholds and does not cause unwanted nonlinear effects. The depletion of the cooling pump power is taken into account in these simulations. The $Yb^{3+}$ ion distribution is a function of the value of the pump, $P_p$, and amplified signal, $P_s$, powers participating in the SRS process. Whilst $Yb^{3+}$ ion distribution in the core of the fiber remains unchanged, any changes in the value of the amplified signal will cause some deviation in the temperature of the fiber from room temperature. Fig. 5 illustrates the deviation of the temperature of the fiber caused by the change in the value of the input amplified signal. The input amplified signals in Fig. 5 are $P_s(0) = 4$ W, $P_s(0) = 4.5$ W, and $P_s(0) = 5.3$ W. The concentration of $Yb^{3+}$ ions in the fiber considered in Fig.5 has been calculated to provide athermal operation of the amplifier, when the input amplified signal $P_s(0) = 5$ W. As one can see in Fig. 5, the changes in the temperature of the fiber caused by the changes in the value of the input amplified signal are inconsequential in comparison with the value of the maximum of the deviation in the temperature of an uncooled fiber. Indeed the change in the the amplified signal of ~20% causes the change in the temperature of the fiber ~ ±10 K, which is an order of magnitude smaller than the peak temperature of the fiber without cooling. This small change in the temperature of the amplifier can be partially compensated for by altering the cooling power.

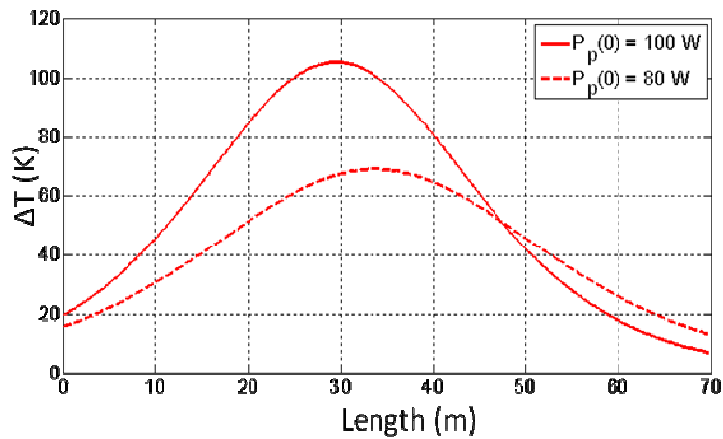

Fig.2. Deviation in the temperature of the fiber ($\Delta T$) along its length for different values of the input pump powers ($P_p(0)$). $P_s(0) = 5$W.

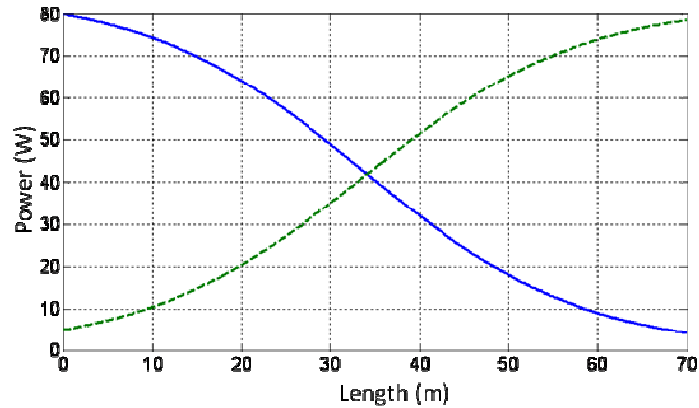

Fig.3. Pump power depletion and signal power amplification along the length of the Raman fiber amplifier.

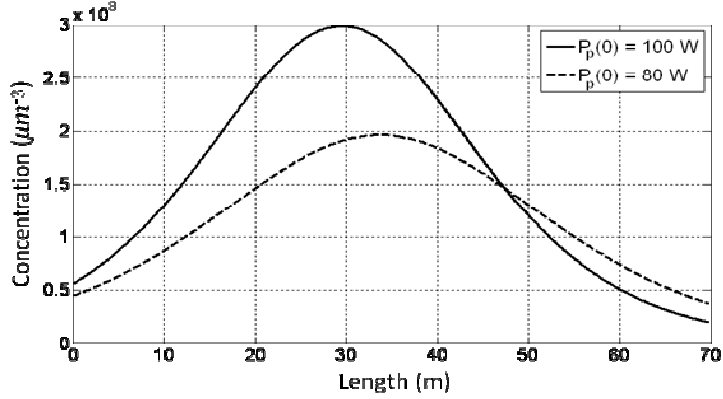

Fig.4. Concentrations of the $Yb^{3+}$ ions in the core of the fiber as functions of the length of the Raman fiber amplifier for different values of the input pump powers, $P_p(0)$. $P_s(0) = 5$W and $P_p^{cool}(0) = 2.2$ W.

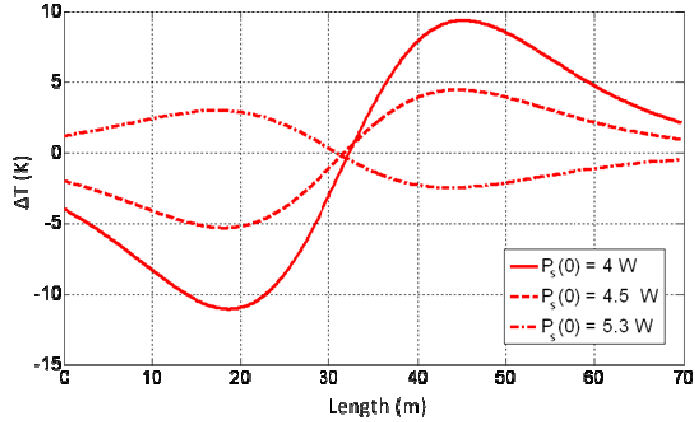

Fig.5. Deviation in the temperature of the fiber ($\Delta T$) along its length for different values of the input signal powers ($P_s(0)$). $P_p^{cool}(0) = 2.2$ W.

The main difficulty encountered for a feasible scheme is the lack of a technique permitting the longitudinal $Yb^{3+}$ distribution in the core of the fiber replicating the curve illustrated in Fig. 4. To overcome this drawback we suggest replacing the ideal fiber cooler design with sections of spliced uniformly doped fibers (Fig.6). Optimizing lengths and concentrations of the ions in the sections of a sequence one can get a distribution of the temperature along the length of the fiber, which oscillates about room temperature with amplitude, which does not exceed several K. These deviations in the temperature are of order value smaller than the value of the peak temperature of the fiber (Fig.2), when the cooling is not applied.

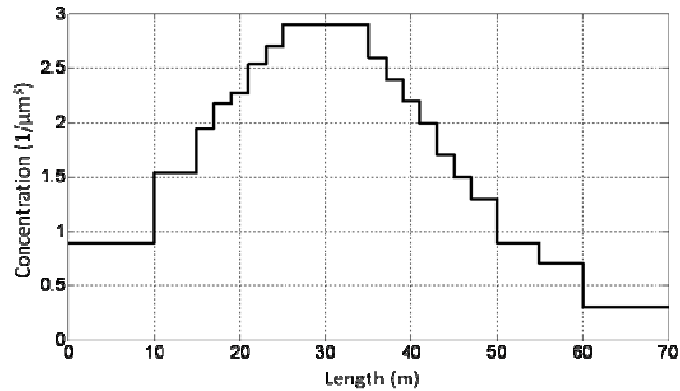

Fig.6. Concentrations of the $Yb^{3+}$ ions in the core of the fiber as functions of the length of the Raman fiber amplifier. $P_p(0) = 100$ W, $P_s(0) = 5$ W. $P_p^{cool}(0) = 2.2$ W.

## 4. Conclusions

In summary, a theoretical scheme of a novel laser cooled high power Raman optical fiber amplifier has been presented. This new mode of amplifier operation should result in little or no heat generation within the active medium. For the first time to our knowledge, it is suggested to integrate the optical cooler consisting of a laser pumped $Yb^{3+}$ ions in the core of the fiber amplifier, where amplification process takes place. The pump and amplified signals of the Raman fiber amplifier propagating at the wavelength $\lambda_p$ = 2.5 $\mu m$ and $\lambda_s$ = 2.575 $\mu m$ (corresponding to the first Stokes wavelengths), respectively, do not interact with $Yb^{3+}$ ions leading to cooling. The concept of integrating a cooler into the active part of the fiber amplifier or laser is radically novel and indeed very promising. Contrary to traditional surface cooling techniques, which strain the amplifier or laser medium and cause distortion of optical waves, the cooler integrated in an active medium can compensate for the heat generated in the amplification process *locally* in the volume of all active media. This mode of cooling is free from the unwanted temperature gradient in the active medium of the device. The integrated cooler also makes the scheme compact, and can be applied to chemically unstable fibers, in which air cooling is not sufficient and water cooling undesirable, such as in ZBLAN glass fibers. The system proposed should be applicable to many other laser and amplifier systems.